# Secure and Privacy- Aware Searching in Peer-to-Peer Networks

Jaydip Sen

TCS Innovation Labs, Tata Consultancy Services Ltd.,
Bengal Intelligent Park, Salt Lake Electronics Complex, Kolkata – 700091, India
`Jaydip.Sen@tcs.com`

**Abstract.** The existing peer-to-peer networks have several problems such as fake content distribution, free riding, white-washing and poor search scalability, lack of a robust trust model and absence of user privacy protection mechanism. Although, several trust management and semantic community-based mechanisms for combating free riding and distribution of malicious contents have been proposed by some researchers, most of these schemes lack scalability due to their high computational, communication and storage overhead. This paper presents a robust trust management scheme for P2P networks that utilizes topology adaptation by constructing an overlay of trusted peers where the neighbors are selected based on their trust ratings and content similarities. While increasing the search efficiency by intelligently exploiting the formation of semantic community structures by topology adaptation among the trustworthy peers, the scheme provides the users a very high level of privacy protection of their usage and consumption patterns of network resources. Simulation results demonstrate that the proposed scheme provides efficient searching to good peers while penalizing the malicious peers by increasing their search times as the network topology stabilizes.

## 1 Introduction

The term *peer-to-peer* (P2P) system encompasses a broad set of distributed applications which allow sharing of computer resources by direct exchange between systems. The goal of a P2P system is to aggregate resources available at the edge of Internet and to share it cooperatively among users. Specially, the file sharing P2P systems have become popular as a new paradigm for information exchange among large number of users in Internet. They are more robust, scalable, fault tolerant and offer better availability of resources. Depending on the presence of central server, P2P system can be classified as centralized or decentralized [1]. In decentralized architecture, both resource discovery and download are distributed. Decentralized P2P application may be further classified as structured or unstructured network. In structured network, there is a restriction on the placement of content and network topology. In unstructured P2P network, however, placement of content is unrelated to topology. Unstructured P2P networks perform better than their structured counterpart in dynamic environment. However, they need efficient search mechanisms and suffer from fake content distribution, free riding (peers who do not share, but consume resources), whitewashing (peers who leave and rejoin the system in order to avoid penalties) and search scalability problems. Open and anonymous nature of P2P applications lead to complete lack of accountability of the content a peer puts in the network. The malicious peers often use these networks to do content poisoning and to distribute harmful programs such as Trojan Horses and viruses [2]. *Distributed reputation based trust management systems* have been proposed to provide protection against malicious content distribution [3]. The main drawbacks of these schemes are their high message exchange overheads and their susceptibility to misrepresentation. Guo et al. have proposed trust-aware adaptive P2P topology to control free-riders and malicious peers [4]. In [5], and [6] topology adaptation is used to reduce inauthentic file download. However, these schemes do not work well in unstructured networks. Poor search scalability is another problem. Traditional mechanisms such as controlled flooding, random walker and topology evolution all lack scalability. Zhuge et al. have proposed trust-based probabilistic search algorithm called *P-walk* to improve search efficiency and to reduce unnecessary traffic in P2P networks [7]. In P-walk, neighboring peers assign trust scores to each other. During routing, peers preferentially forward queries to the highly ranked neighbors. However, its performance in large-scale unstructured network is questionable. To combat free riders, various trust-based incentive mechanisms are presented in [8]. Most of these mechanisms, however, involve large overhead of computations.

To combat the problem of inauthentic downloads as well as to improve search scalability while protecting the privacy of the users, this paper proposes an adaptive trust-aware algorithm that is robust and scalable. This work is an extension of our already published scheme which is based on construction an overlay of trusted peers where neighbors are selected based on their trust ratings and content similarities [9]. It increases search efficiency by taking advantage of implicit semantic community structures formed as a result of topology adaptation since most of the queries are resolved within the community [9]. However, the novel contribution of the current work is that it combines the functionalities of a robust trust management model and the semantic community formation that also ensures user privacy is protected. While the trust management scheme segregates honest peers from malicious peers, based on both first-hand and second-hand information, the semantic community formation allows topology adaptation to form cluster of peers which share similar contents. The formation of the semantic communities also enables the scheme to form a neighborhood of trust which is utilized to protect user privacy in the network.

The rest of the paper is organized as follows. Section 2 discusses some related work. Section 3 presents the proposed algorithm for secure and privacy-aware searching. For the benefit of the readers, we present the entire algorithm including the one presented in [9]. As mentioned in the previous paragraph, the scheme presented in this paper has a robust trust management model and privacy preserving module which were not present in the scheme described in [9]. Section 4 introduces various metrics to measure performance of the proposed algorithm, and presents the simulation results. Section 5 concludes the paper while highlighting some future scope of work.

## 2   Related Work

In [10], a searching mechanism is proposed that is based on discovery of trust paths among the peers in a peer-to-peer network. A global trust model based on distance-weighted recommendations has been proposed in [11] to quantify and evaluate the peers in a peer-to-peer network. In [5], a protocol named *adaptive peer-to-peer technologies* (APT) for the formation of adaptive topologies has been proposed to reduce spurious file download and free riding, where a peer connects to those peers from whom it is most likely to download satisfactory content. It adds or removes neighbors based on *local trust* and *connection trust* which are decided by its transaction history. The scheme follows a defensive strategy for punishment where a peer equally punishes both malicious peers as well as neighbors through which it receives response from malicious peers. This strategy is relaxed in the *reciprocal capacity-based adaptive topology protocol* (RC-ATP), where a peer connects to others which have higher *reciprocal capacity* [6]. Reciprocal capacity is defined based on peers's capacity of providing good files and of recommending source of download. While RC-ATP provides better network connectivity than APT, and reduces the cost of inauthentic downloads, it has a large overhead of topology adaptation.

There are some significant difference between the proposed algorithm and APT and RC-ATP. First, in the proposed scheme, the links in the original overlays are never deleted to avoid network partitioning. Second, the robustness of the proposed protocol in presence of malicious peers is higher than that of APT and RC-ATP protocols as shown in the experimental results. Third, as APT and RC-ATP both use flooding to locate resource, they have poor search scalability. The proposed scheme takes the advantages of semantic communities to improve QoS of search. Fourth, APT and RC-ATP do not employ any robust trust model for security in searching and for user identity and data privacy protection. The central part of the proposed searching mechanism in this paper is a robust trust management model. Finally, unlike APT and RC-ATP, the proposed algorithm scheme punishes malicious peers by blocking query initiated by them.

## 3   The Proposed Secure and Privacy-Aware Searching Algorithm

This section is divided into two parts. In the first part, the various parameters for P2P networks are discussed. In the second part, the proposed algorithm is presented.

### 3.1  The Network Environment

To derive meaningful conclusion from the proposed algorithm, the proposed scheme have been modeled in P2P networks in a realistic fashion. The factors that are taken into consideration are as follows.

(1)*Network topology and load*: The topology of the network plays an important role for the analysis of trust management and search procedure. Following the work in [5][6], the network has been modeled as a *power law graph*. In a power law network, degree distribution of nodes follows power law distribution, i.e. fraction of nodes having degree $L$ is $L^{-k}$ where $k$ is a network dependent constant. Prior to each simulation cycle a fixed fraction of peers chosen randomly is marked as malicious. As the algorithm proceeds, the peers adjust topology locally to connect those peers which have better chance to provide good files in future and drop malicious peers from their neighborhood. The network links are categorized into two types: *connectivity link* and *community link*. The connectivity links are the edges of the original power law network which provide seamless connectivity among the peers. To prevent the network from being fragmented they are never deleted. On the other hand, community links are added probabilistically between the peers who know each other. A community link may be deleted when perceived trustworthiness of a peer falls in the perception of its neighbors. A limit is put on the additional number of edges that a node can acquire to control bandwidth usage and query processing overhead in the network. This increase in network load is measured relative to the initial network degree (corresponding to connectivity edges). Let *final_degree(x)* and *initial_degree(x)* be the initial and final degree of a node *x*. The *relative increase in connectivity* (RIC) is constrained by a parameter known as *edge_limit*.

$$RIC = \frac{final\_degree(x)}{initial\_degree(x)} \leq edge\_limit \qquad (1)$$

(2) *Content distribution*: The dynamics of a P2P network are highly dependent on the volume and variety of files each peer chooses to share. Hence a model reflecting real-world P2P networks is required. It has been observed that peers are in general interested in a subset of the content on the P2P network [12]. Also, the peers are often interested only in files from a few content categories. Among these categories some are more popular than others. It has been shown that Gnutella content distribution follows *zipf distribution* [13]. Keeping this in mind, both content categories and file popularity within each category is modeled with *zipf distribution* with $α = 0.8$.

*Content distribution model*: The content distribution model in [13] is followed for simulation purpose. In this model, each distinct file $f_{c,r}$ is abstractly represented by the tuple $(c, r)$, where $c$ represents the content category to which the file belongs, and $r$ represents its popularity rank within a content category $c$. Let content categories be $C = \{c_1, c_2,…,c_{32}\}$. Each content category is characterized by its *popularity rank*. For example, if $c_1 = 1$, $c_2 = 2$ and $c_3 = 3$, then $c_1$ is more popular than $c_2$ and hence it is more replicated than c2 and so on. Also there are more files in category $c_1$ than $c_2$.

Table 1. Hypothetical content distribution in peer nodes

| **Peers** | **Content categories** |
|---|---|
| $P_1$ | $\{C_1, C_2, C_3\}$ |
| $P_2$ | $\{C_2, C_4, C_6, C_7\}$ |
| $P_3$ | $\{C_2, C_4, C_7, C_8\}$ |
| $P_4$ | $\{C_1, C_2\}$ |
| $P_5$ | $\{C_1, C_5, C_6\}$ |

Each peer randomly chooses between three to six content categories to share files and shares more files in more popular categories. Table 1 shows an illustrative content distribution among 5 peers. The category $c_1$ is more replicated as it is most popular. The *Peer 1* shares files in three categories: $c_1$, $c_2$, $c_3$ where it shares maximum number of files in category $c_1$, followed by category $c_2$ and so on. On the other hand, *Peer 3* shares maximum number of files in category $c_2$ as it's the most popular among the categories chosen by it, followed by $c_4$ and so on.

(3) *Query initiation model*: The authors in [13] suggest that peers usually query for files that exist on the network and are in the content category of their interest. In each cycle of simulation, active peers issue queries. However number of queries a peer issues may vary from peer to peer, modeled by *Poisson* distribution as follows. If *M* is the total number of queries to be issued in each cycle of simulation and *N* is the number of peers present in the network, query rate $\lambda = M/N$ is the mean of the *Poisson* process. The expression $p(\#quries = K) = \frac{e^{-\lambda}\lambda^K}{K!}$ gives the probability that a peer issues *K* queries in a cycle. The probability that a peer issues query for the file $f_{c,r}$ depends on the peer's interest level in category *c* and rank *r* of the file within that category.

(4) *Trust management engine*: A trust management engine is designed which helps a peer to compute trust ratings of other peers from past transactions as well as recommendation from its neighbor. For computation of trust values for the peers, a method similar to the one proposed in [14] is followed. The framework employs a beta distribution for reputation representation, updates and integration. The first-hand information and second-hand (recommendation from neighbors) are combined to compute the reputation value of a peer. The weight assigned by a peer *i* to a second-hand information received from a node *k* is a function of reputation of node *k* as maintained in node *i*. For each peer *j*, a reputation $R_{ij}$ is computed by a neighbor peer *i*. The reputation is embodied in the *Beta* model which has two parameters $\alpha_{ij}$ and $\beta_{ij}$. $\alpha_{ij}$ represents the number of successful transactions (i.e. authentic file downloads) that peer *i* had with peer *j*, and $\beta_{ij}$ represents the number of unsuccessful transactions (i.e., unauthentic file downloads). The reputation of peer *j* maintained by peer *i* is computed using (2):

$$R_{ij} = Beta(\alpha_{ij} + 1, \beta_{ij} + 1) \tag{2}$$

The trust metric of a peer is the expected value of its reputation and is given by (3):

$$T_{ij} = E(R_{ij}) = E(Beta(\alpha_{ij} + 1, \beta_{ij} + 1)) = \frac{\alpha_{ij} + 1}{\alpha_{ij} + \beta_{ij} + 2} \tag{3}$$

The second-hand information is presented to peer *i* by its neighbor peer *k*. The peer *i* receives the reputation $R_{kj}$ of peer *j* from peer *k*, in the form of the two parameters $\alpha_{kj}$ and $\beta_{kj}$. After receiving this new information the peer combines it with its current assessment $R_{ij}$ to obtain a new reputation $R_{ij}^{new}$ as shown in (4):

$$R_{ij}^{new} = Beta(\alpha_{ij}^{new}, \beta_{ij}^{new}) \tag{4}$$

In (4) the values of $\alpha_{ij}^{new}$ and $\beta_{ij}^{new}$ are given by (5) and (6) as follows:

$$\alpha_{ij}^{new} = \alpha_{ij} + \frac{2\alpha_{ik}\alpha_{kj}}{(\beta_{ik}+2)(\alpha_{kj}+\beta_{kj}+2)(2\alpha_{ik})} \quad (5)$$

$$\beta_{ij}^{new} = \beta_{ij} + \frac{2\alpha_{ik}\beta_{kj}}{(\beta_{ik}+2)(\alpha_{kj}+\beta_{kj}+2)(2\alpha_{ik})} \quad (6)$$

The proposed trust model gives more weight to recent observations, which is used for updating the reputation value suing direct observation. For updating the reputation value using the second-hand information, *Dempster-Shafer* theory [15] and the *belief discouting model* [16] are used. The reputation of a recommending peer is automatically taken into account while computing the reputation of the reported peer. This eliminates the need of a separate deviation test. As mentioned earlier in this section, the trust value of a peer is computed as the statistical expected value of its reputation. The trust value of a peer lies in the interval [0, 1]. Peer *i* considers peer *j* as trustworthy if $S_{ij} \geq 0.5$, and malicious if $S_{ij} < 0.5$.

(5) *Identity of the peers*: Each peer generates a 1024 bit public/private RSA key pair. The public key serves as the identity of the peer. The identities are persistent and they enable two peers that have exchanged keys to locate and connect to one another whenever the peers are online. In addition, a *distributed hash table* (DHT) is maintained that lists the transient IP-addresses and port numbers for all peers for all applications running of the peers. DHT entries for the peer *i* are signed by *i* and encrypted with its public key. Each entry is indexed by a 20 byte randomly generated shared secret, which is agreed upon during the first successful connection between two peers. Each peer's location in the DHT is independent of its identity and is determined by hashing the client's current IP address and DHT port. This inhibits systematic monitoring of targeted regions of the DHT key space since the region foe which each peer is responsible is determined by that peer's network address and port.

(6) *Node churning model*: In P2P networks, a large number of peers may join and leave at any time. This activity is termed as node *churning*. To simulate node churning, prior to each *generation* (a set of consecutive searches), a fixed percentage of nodes are chosen randomly as inactive. These peers neither initiate nor respond to a query in that generation and join the system latter with their LRU structure cleared. Since in a real world network, even in presence of churning, the approximate distribution of content categories and files remain constant, content of nodes undergoing churn is exchanged which in effect assigns each of them new content as well as keeps content distribution model of the network unchanged.

(7) *Threat model*: Malicious peers adopt various strategies (threat model) to conceal their behavior and disrupt system activity. Two threat models are considered in the proposed scheme. The peers who share good quality files enjoy better topological due to topology adaptation. In *threat model A*, malicious peers attempt to circumvent this by providing good file occasionally with probability, known as *degree of deception* to lure other peers to form communities with them. In *threat model B*, a group of malicious peer joins to the system and provides good files until their connectivity reaches to *edge limit*, and then start spreading fake content in the network.

### 3.2 The proposed search algorithm

The network learns trust information through the search and updates trust information and adapts topology based on the outcome of the search. The following criteria are kept in mind while designing the algorithm: (1) It should improve search efficiency as well as search quality (authentic file download). (2) It should have minimal overhead in terms of computation, storage and message passing. (3) It should provide incentive to share large number of high quality files. (4) It should be self policing in the sense that a peer can adjust search strategy based on local estimate of network connectivity. (5) It should be able to protect the privacy of the users. Major steps of the algorithm are: (i) search, (ii) trust computing and verification, and (iii) topology adaptation. Each of these steps is discussed in the following.

#### 3.2.1 Search

A *time to live* (TTL) bound search is used. At each hop, query is forwarded to a subset of neighbors, the number of neighbors is decided based on the local estimate connectivity. This connectivity index for peer $x$ is denoted as $Prob_{com}(x)$ and is given by (7):

$$Prob_{com}(x) = \frac{degree(x) - initial\_degree(x)}{initial\_degree(x).(edge\_limit - 1)} \quad (7)$$

When $Prob_{com}$ for a node is low, the peer has the capacity to accept new community edges and expand community structures. Higher the value of $Prob_{com}$, lesser the neighbors choose to disseminate queries. As simulation proceeds, connectivity of good nodes increases and reaches a maximum value. At this time, the peers focus on directing queries to appropriate community which may host the specific file rather than expanding communities. For example, if peer $i$ can contact at most 10 neighbors and $Prob_{com}$ of $j$ is 0.6, it forwards query to: 10 x (1 – 0.6) = 4 neighbors only. The search strategy modified from initial *TTL limited BFS* to *directed DFS* with the restructuring of the network. The search process has two steps– *query initiation* and *query forward*. These steps are described in the following.

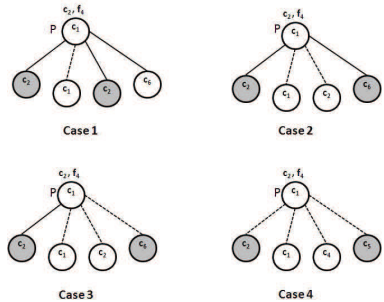

Fig. 1. Neighbor selection at *P* for query string ($c_2$, $f_4$). Community edges and connectivity edges are drawn with solid and dotted lines respectively. Nodes that dispatch query are shaded.

*Query initiation*: The initiating peer forms a query packet containing the name of the file (*c, r*) and forwards it to a certain fraction of neighbors along with $Prob_{com}$ and

TTL value. The query is disseminated using the following *neighbor selection rule*. The neighbors are ranked based on both trustworthiness and the similarity of interest. Preference is given to the trusted neighbors sharing similar contents. Among the trusted neighbors, community members having content matched to the query are preferred. When there are insufficient community links, query is forwarded through connectivity links also. The various cases of neighbor selection are illustrated in Fig. 1. It is assumed that in each case only two neighbors are selected. When the query ($c_2$, $f_4$) reaches node *P*, following cases may occur. In first case, *P* has adequate community neighbors sharing file in category $c_2$, hence they are chosen. In *Case 2*, there is insufficient number of community neighbors sharing file in the requested category, the community neighbors sharing $c_2$ and $c_6$ preferred to the connectivity neighbor $c_2$ to forward query. In *Case 3*, only one community neighbor who share file is $c_2$. Hence it is chosen. From the remaining connectivity neighbors, most trusted $c_6$ is selected. In *Case 4*, only connectivity neighbor are there, assuming all of them at the same trust level, the matching neighbor $c_2$ is chosen and from the rest $c_5$ is selected randomly. The pseudo-code for the *query initiation algorithm* is presented in *Algorithm 1*.

**Algorithm 1** *Query Initiation*
1: Form a query packet containing tuple (*c*,*r*) and TTL.
2: Calculate number of neighbors to whom to forward query using $Prob_{com}$.
3: Select neighbors using *Neighbor Selection Rule*.
4: Forward query packets to the neighbors selected.

When a query reaches to peer *i* from peer *j*, following actions are performed by the peer *i*.

*Query forward*: (i) *Check trust level of peer j*: Peer *i* checks trust rating of peer *j* through *check trust rating* algorithm (explained later). Accordingly decision regarding further propagation of the query is taken. (ii) *Check the availability of file*: If the requested file is found, response is sent to peer *j*. If TTL value has not expired, the following steps are executed. (iii) *Calculate the number of messages to be sent*: It is calculated based on the value of $Prob_{com}$. (iv) *Choose neighbors*: Neighbors are chosen in using *neighbor selection rule*. The search process is shown in Fig. 2. It is assumed that the query is forwarded at each hop to two neighbors. The matching community links are preferred over connectivity links to dispatch query. Peer *1* initiates query and forwards it to two community neighbors *3* and *4*. The query reaches peer *8* via peer *4*. However, peer *8* knows that peer *4* is malicious from previous transactions. Hence it blocks the query. The query forwarded by peer *5* is also blocked by peer *10* and *11* as both of them know that peer *5* is malicious. The query is matched at four peers: *4, 6, 9* and *13*. The pseudo-code for the query forward algorithm is presented in Algorithm 2. The search process is shown in Fig. 2.

*Topology Adaptation*: Responses are sorted by the initiating peer *i* based on the reputation of resource providers and peer having highest reputation is selected as source of download. The requesting peer checks the authenticity of downloaded file. If the file is found to be fake, peer *i* attempts to download from other sources until it finds the authentic resource or no more sources exist and updates the trust rating and possibly adapts topology after failed or successful download, to bring trusted peers to its neighborhood and to drop malicious peers from its community. The restructuring of network is controlled by a parameter known as *degree of rewiring* which is the probability with which a link is formed between two peers. This parameter allows trust information slowly to be propagated as happens in real network. Topology adaptation consists of the following operations: (i) *link deletion*: Peer $i$ deletes the existing community link with peer $j$ if it finds peer $j$ as malicious. (ii) *link addition*: Peer $i$ probabilistically forms community link with peer $j$ if resource is found to be authentic. If $RIC \leq edge\_limit$, for both peers $i$ and $j$, only then an edge can be added subject to the approval of resource provider $j$. If peer $j$ finds that peer $i$ is malicious, it doesn't approve the link.

**Algorithm 2** *Query Forward*
1: Input: A query packet containing tuple $(c,r)$ from peer $j$.
2: $available(c,r)$: A boolean function returns true if $(c,r)$ is locally available.
3: Calculate trust score of peer $j$.
4: **if** $score \leq 0$ **then**
5:    Exit.
6: **else**
7:    **if** $available(c,r)$ **then**
8:       Send response to peer $j$.
9:    **end if**
10:    $TTL \leftarrow TTL - 1$
11:    **if** $TTL \geq 0$ **then**
12:       Calculate number of neighbors using the formula $(1 - Prob_{com}) \times N$ where N be total number of edges.
13:       Select neighbors using *Neighbor Selection Rule*.
14:       Forward query packets to the neighbors selected.
15:    **end if**
16: **end if**

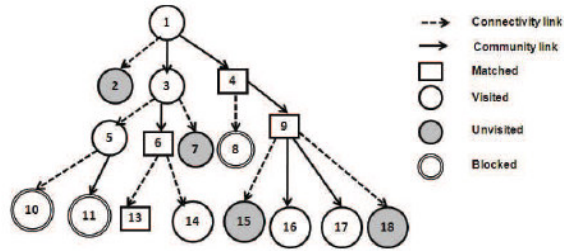

Fig. 2. The breadth first search (BFS) tree for the search procedure initiated by peer 1.

In the example shown in Fig. 3, peer 1 downloads the file from peer 4 and finds that the file is spurious. It reduces the trust score of peer 4 and deletes the community link 1-4. It then downloads the file from peer 6 and gets an authentic file. Peer 1 now sends a request to peer 6, and the latter grants the request after consulting its LRU and the community edge 1-6 is added. The malicious peer 4 loses one community link and peer 6 gains one community edge. However, the network still remains connected by connectivity edges, shown in dotted lines. Algorithm 3 presents the pseudo-code for the topology adaptation process.

*Check trust rating*: Trust rating is used at various stage of the algorithm to make decision about the possible source for download, to stop a query forwarded from a malicious node and to adapt topology. A *least recently used* (LRU) data structure is used at each peer to keep track of *32* most recent peers it has interacted with. When no transaction history is available, a peer seeks recommendation from its neighbors using *trust query*. When peer *i* doesn't have trust score of peer *j* in its LRU history, it first seeks recommendation about *j* from all of its community neighbors. If none of its community neighbors possesses any information about *j*, peer *i* initiates directed DFS search. The trust computation model has been presented in Section 3.1.

**Algorithm 3** *Topology Adaptation* (executed by initiator $i$)
1: A set $S$ containing responses.
2: Sort S based on the trust score of resource provider.
3: **repeat**
4:    Download file from the source j having highest trust score among the rest.
5:    Check file status and update trust score of source.
6:    *Rewire network(i,j)*.
7: **until** file is authentic or no more source exist in S

**Algorithm 4** *Rewire Network*
1: $i$ is initiator, $j$ is resource provider.
2: **if** source $j$ is malicious **then**
3:    Delete existing community edge $(i,j)$.
4: **else**
5:    $R = rand(0, 1)$
6:    **if** $rand \leq Prob_{comm}$ **then**
7:      **if** *Approve Link (i,j)* **then**
8:        Form community edge (i,j) .
9:      **end if**
10:   **end if**
11: **end if**

**Algorithm 5** *Approve Link*
1: $i$ is initiator, $j$ is resource provider.
2: Return: Boolean value.
3: **if** $RIC(i) or RIC(j) \geq edge\_limit$ **then**
4:    return 0.
5: **else if** $score(i) < 0$ **then**
6:    return 0.
7: **else**
8:    return 1.
9: **end if**

### 3.2.2 Privacy-preservation in searching

The trust-based searching scheme described above does not guarantee any privacy requirement of the requester (i.e. the initiator of the query). For protecting the privacy of the user, several enhancement of the algorithm are proposed. Following cases are identified for privacy preservation.

(a) *Identity of the requesting peer is to be protected*: In this case, as shown in Fig 4. instead of sending the request straightway to the supplier peer, the requesting peer asks one of its trusted peers (which may or may not be its neighbor) to look up the data on its behalf. Once the query propagation module successfully identifies the possible supplier of the resource, the trusted peer serves as a proxy to deliver the data to the requester node. Other peers including the supplier of the resource will not be able to know the real requester. Hence, the requester's privacy is protected. Since the requestor's identity is only known to its trusted peer, the strength of privacy is dependent on the effort required to compromise the trusted peer. As mentioned in Section 3.1, the message communicated the peers are encrypted by 1024 bit RSA key, which is a provably secure algorithm. Hence, the privacy of the requester is highly protected.

(b) *Protecting the data handle*: To improve the achieved privacy level, the data handle may not be put in the request at the beginning. When a requester initiates the request, it computes the hash value of the handle and reveals only a part of the hash result in the request sent to its trusted peer. The steps 1 and 2 in Fig 5 represent these activities. Each peer receiving the request compares the revealed partial hash to hash codes of the data handles that it holds. Depending on the length of the revealed part, the receiving peer may find multiple matches. This does not, however, imply that the peer has the requested data. Thus this peer will provide a candidate set, along with a certificate of its public key, to the requester. If the matched set is not empty, the peer will construct a Bloom filter [17] based on the left parts of the matched has codes, and send it back to the trusted peer. The trusted peer forwards it back to the requester. These are represented by the steps 3 and 4 in Fig. 5. Examining the filters, the requester can eliminate from the candidate data supplier list all peers that do not have the required data. It then encrypts the complete request with the supplier's public key and gets the requested data with the help from its trusted node. The steps 5, 6, 7 and 8 in Fig 5 represent these activities. By adjusting the length of the revealed hash code, the requestor can control the number of eliminated peers. The level of privacy is much improved than the previous case since the malicious peers need to both compromise the trusted node and break the Bloom filter and has function.

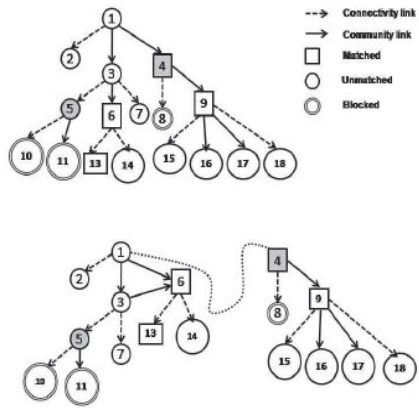

Fig. 3. Topology adaptation based on outcome of the search in Figure 2. Malicious nodes are shaded in gray color.

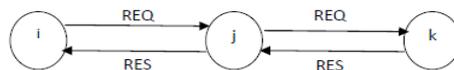

Fig. 4. Identity protection of the requesting peer i from the supplier peer k by use of trusted peer j. REQ and RES are the request and response messages respectively.

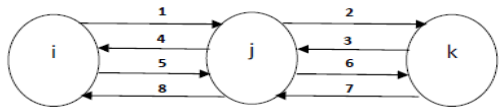

Fig. 5. Protecting data handle using trusted node. Peer *i* and *k* are the requester and the supplier peer. Peer *j* is the trusted peer of the requester peer *i*.

(c) *Hiding the data content*: Although the privacy-preservation level has been improved during the look-up phase using the previous two schemes, the privacy of the requester will be compromised if the trusted node can see the data content when it relays the packets for the requester. To improve privacy level and prevent eavesdropping, we can encrypt the data handle and the data content. If the identity of the supplier is known to the requester, it can encrypt the request using the supplier's public key. The public key of the requester cannot be used because the certificate will reveal its identity. The problem is solved in the following manner. The requester generates a symmetric key and encrypts it using a supplier's public key. Only the supplier can recover the key and use it to encrypt data. To prevent the trusted node of the requester from conducting a man-in-the-middle attack, the trusted node is required to sign the packet. This provides a non-repudiation evidence, and shows that the packet is not generated by the trusted node itself. The privacy level has been improved since now the malicious nodes need to break the encryption keys as well.

## 4 Performance Evaluation

To analyze the performance of the proposed algorithm, several metrics are defined. Due to constraints of space, the performances of the algorithm for some of the metrics are presented in this paper. Additional experimental results may be found in [9].

(a) *Attempt ratio* (AR): A peer keeps on downloading files from various sources based on their trust rating till it gets the authentic file. AR is the probability that the authentic file is downloaded in the first attempt. A high value of AR is desirable.

(b) *Effective attempt ratio* (EAR): It measures the cost of downloading an authentic file by a good peer in comparison to the cost incurred by a malicious peer.

If $P(i)$ be the total number of attempts made by the peer $i$ to download an authentic file, EAR is given by (8):

$$EAR = (\frac{1}{M} \sum_{i=1}^{M} \frac{1}{P(i)} - \frac{1}{N} \sum_{j=1}^{N} \frac{1}{P(j)}) \times 100 \qquad (8)$$

In (8), $M$ and $N$ are the number of malicious and good peers issuing queries in a particular generation. For example, EAR = 50 implies that if a good peer needs one attempt to download an authentic file, a malicious peer will need two attempts.

(c) *Query miss ratio* (QMR): Since the formation of semantic communities takes some time, there will be a high rate of query misses in the first few generations of search. However, as the algorithm executes, the rate of query miss is expected to fall for the good peers. QMR is defined as the ratio of the number of search failures to the total number of searches in a generation.

(d) *Relative increase in connectivity* (RIC): After a successful download, a requesting peer attempts to establish a community edge with the resource provider, if approved by the latter. This ensures that peers which provide good community services are rewarded by having increasing number of community neighbors. The metric RIC measures the number of community neighbors a peer gains with respect to its connectivity neighbors in the initial network topology. If $D_{init}(i)$ and $D_{final}(i)$ are the initial and final degrees of the peer $i$, and $N$ is the number of peers, then RIC for peer $i$ is computed using (9). As discussed in [9], the connectivity of good peers increases significantly over time.

$$RIC = \frac{1}{N} \sum_i \frac{D_{final}(i)}{D_{init}(i)} \quad (9)$$

(e) *Closeness centrality* (CC): Since the topology adaptation effectively brings the good peers closer to each other, the length of the shortest path between a pair of good decreases. This intrinsic incentive for sharing authentic files is measured by the metric CC. The peers with higher CC values are topologically better positioned. If $P_{ij}$ is the length of the shortest path between peer $i$ and peer $j$ through the community edges and if $V$ denotes the set of peers, then CC for peer $i$ is given by (10).

$$CC_i = \frac{1}{\sum_{j \in V} P_{ij}} \quad (10)$$

(f) *Clustering coefficient* (CLC): It gives an indication about how well the network forms cliques and plays an important role in choosing the TTL value. Higher the value of CLC, lower TTL value can be used. If $K_i$ be the number of community neighbors of peer $i$, then clustering coefficient (CLS) of peer $i$ is defined as:

$$CLC(i) = \frac{2 \times E_i}{K_i \times (K_i - 1)} \quad (11)$$

In (11), $E_i$ is the actual number of community edges between the $K_i$ neighbors. CLC of the network is taken as the average value of all CLC($i$)s.

(f) *Trust query propagation overhead* (TQPO): The peers build trust and reputation information both by collection of first-hand and second-hand information. Trust query message is propagated when trust information about a peer is not available locally. A trust query message involves one DFS round without backtracking. The overhead incurred due to trust query propagation is measured by the metric called *trust query propagation overhead* (TQPO). TQPO is defined as the total number of distinct DFS search attempts per generation. It may be noted that a trust query may be initiated multiple times for a single file search operation: to select a trusted neighbor or to approve a community link.

(g) *Largest connected component* (LCC): The community edges connect nodes with similar content interest and having mutual trust on each other. If we consider the peers which share a particular category of contents, then the community edges form a trust-aware community overlay. However, it will be highly probable that the trust-aware overly graph will be a disconnected graph. LCC is the largest connected component of this disconnected overlay graph. LCC of the network can be taken as a measure of the goodness of the community structure since it signifies how strongly the peers with similar contents and interests are connected with each other. It is expressed in terms of the percentage of nodes of a particular category that lies within the LCC.

A discrete time simulator written in C is used for simulation. In simulation, 6000 peer nodes, 18000 connectivity edges, 32 content categories are chosen. The degree of deception and the degree of rewiring are taken as 0.1 and 0.3 respectively. The value of the edge_limit is taken as 0.3. The TTL values for BFS and DFS are taken as 5s and 10 s respectively. The discrete time simulator simulates the algorithm repeatedly on the power law network and outputs all the metrics averaged over generations. Barabasi-Albert generator is used to generate initial power law graphs with 6000 nodes and approximately 18000 edges. The number of search per generation is taken as 5000 while the number of generations per cycle of simulation is 100.

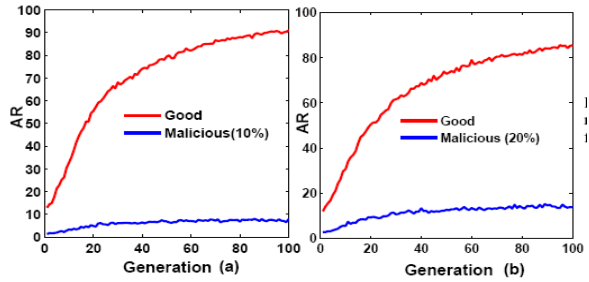

Fig. 6. AR vs. percentage of malicious nodes. In (a) 10% , in (b) 20% nodes are malicious

To check the robustness of the algorithm against attack from malicious peers, the percentage of malicious peers is gradually increased. Fig. 6 illustrates the cost incurred by each type of peers to download authentic files. As the percentage of malicious peers is increased, cost incurred by malicious peers to download authentic files decreases while that of good peers increases.

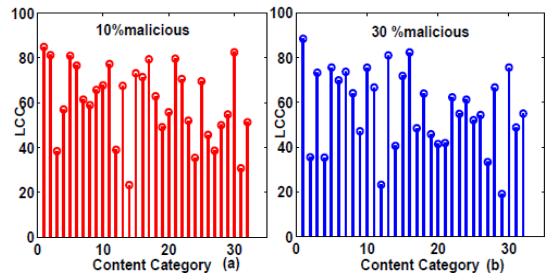

Fig. 7. Largest connected components (LCC) for different content categories

Fig. 7 depicts the size of LCC for each of the 32 content categories. It may be observed that the average size of LCC for all content categories remains even if the percentage of malicious peers increases. This clearly shows that the community formation among the honest peers is not affected by the presence of malicious nodes.

Fig. 8 presents how the *closeness centrality* (CC) of good and malicious peers varies in the community topology. In computation of CC, only the community edges have been considered. It may be observed that the steady state value of CC for honest peers is around 0.12. However, for the malicious peers, the CC value is found to lie between 0.03 to 0.07. This demonstrates how effectively the malicious peers are driven to the fringe of the network while the good peers are rewarded.

Higher values of CC also indicate that good peers have smaller average shortest path length between them. In the simulation, the diameter of the initial network is taken as 5. At the end of one simulation run, if there is no path between a pair of peers using community edges, then the length of the shortest path between that pair is assumed to be arbitrarily long, say 15 (used in Fig. 9). As shown in Fig. 9, the *average shortest path distance* (ASPD) decreases form the initial value of 15 for both honest and malicious nodes. However, the rate and the extent of decrease for honest peers are much higher due to the formation of semantic communities. For malicious peers, after an initial fall, the value of ASPD increases consistently and finally almost reaches the maximum value of 15. On the other hand, the average value of ASPD for honest peers is observed to be around 6. Since the honest nodes are connected with shorter paths, the query propagations and their responses will also be faster among these nodes.

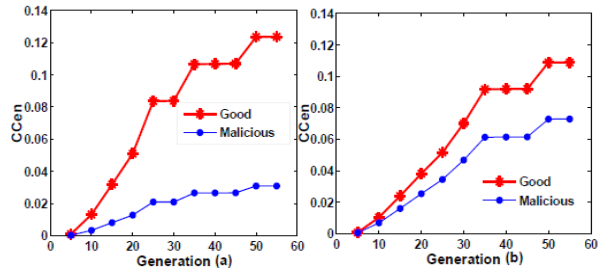

Fig. 8. Closeness centrality for various percentages of malicious nodes. In (a) 20% and (b) 40% nodes are malicious.

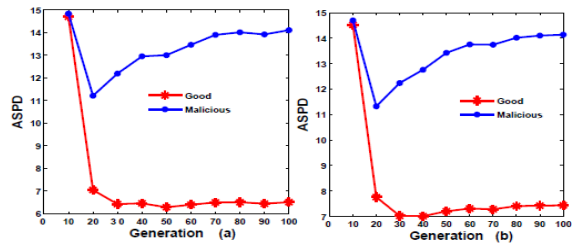

Fig. 9. Avg. shortest path distance vs generations of search at the step of ten for various percentages of malicious peers. In (a) 30% and in (b) 40% nodes are malicious.

Fig. 10 shows that as the topology of the network matures, the steady state value of *trust query propagation overhead* (TQPO) attains a quite low value – less than 10 when 10% of the peers in the network are malicious. Even when the network has 40% of its peers malicious, TQPO gradually decreases and is less than 20 within 100 generations. Hence trust propagation has little impact on the system overhead since the trust information gets embedded in trust-aware overlay topology. Moreover, the storage requirement is low due to usage of the LRU structure.

Fig. 11 shows *clustering coefficient* (CLC) for each type of peers. Since community edges are added based on the download history and peers having good reputation gain more community edges, clustering coefficient (CLC) is high for good peers. This leads to triangle formation in the communities. To counter this phenomenon, the search strategy adapts itself from BFS to DFS to minimize redundant message flows in the network. Since edges are added based on the download history and similarity of inter-est, community of peers are formed which are connected to other community by hub peers having interest in multiple content categories. This leads to lower ASPD for good peers.

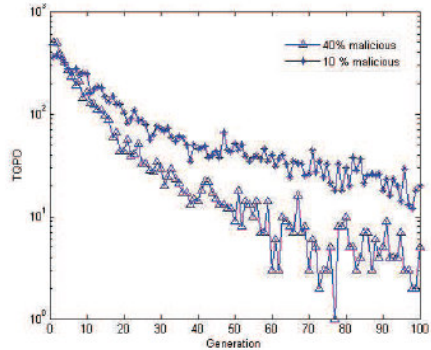

Fig. 10. Overhead of trust query propagation for 10% and 20% malicious peers in the network

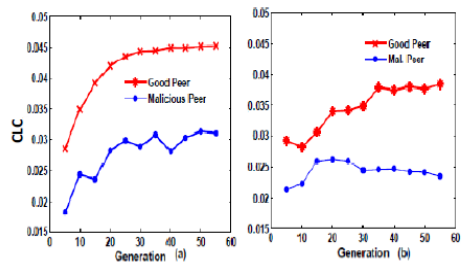

Figure 11. Clustering coefficients for different percentages of malicious peers. In (a) 20% and in (b) 40% of the peers are malicious.

## 5 Conclusion

In this paper, a search mechanism is proposed that solves multiple problems in peer-to-peer network e.g., inauthentic download, poor search scalability, combating free riders and protecting user privacy. It is shown that by topology adaptation, and robust trust management, it is possible to isolate the malicious peers while providing topologically advantageous positions to the good peers so that good peers get faster and authentic responses to their queries. Simulation results have demonstrated that the protocol is robust in presence of a large percentage of malicious peers. Analysis of message overhead of the privacy module constitutes a future plan of work.

## References

1. Risson, J., Moors, T.: Survey of Research Towards Robust Peer-to-Peer Networks. Computer Networks, 50(7), 3485-3521. (2006)

2. Schafer, J., Malinks, K., Hanacek, P.: Peer-to-Peer Networks Security. In: Proc. of the 3$^{rd}$ Int. Conf. on Internet Monitoring and Protection (ICIMP), pp. 74-79. (2008).
3. Abdul-Rahman, A., Hailes, S.: A Distributed Trust Model. In: Proc. of the Workshop on New Security Paradigms, pp. 48-60. (1997)
4. Guo, L., Yang, S., Guo, L., Shen, K., Lu, W.: Trust-Aware Adaptive P2P Overlay Topology Based on Super-Peer-Partition. In: Proc. of the 6$^{th}$ Int. Conf. on Grid and Cooperative Computing, pp. 117-124. (2007)
5. Condie, T., Kamvar, S.D., Garcia-Molina, H.: Adaptive Peer-to-Peer Topologies. In: Proc. of the 4$^{th}$ Int. Conf. on Peer-to-Peer Computing (P2P'04), pp. 53-62. (2004)
6. Tain, H., Zou, S., Wang, W., Cheng, S.: Constructing Efficient Peer-to-Peer Overlay Topologies by Adaptive Connection Establishment. Computer Communication, 29(17), 3567-3579. (2006)
7. Zhuge, H., Chen, X., Sun, X.: Preferential Walk: Towards Efficient and Scalable Search in Unstructured Peer-to-Peer Networks. In: Proc. of the 14$^{th}$ Int. Conf. on World Wide Web (WWW'05), Poster Session, pp. 882-883. (2005)
8. Tang, Y., Wang, H., Dou, W.: Trust Based Incentive in P2P Network. In: Proc. of the IEEE Int. Conf. on E-Commerce Technology for Dynamic E-Business, pp. 302-305. (2004)
9. Sen, J.: A Robust and Fault-Tolerant Distributed Intrusion Detection System. In: Proc. of the 12$^{th}$ International Conference on Information and Communication Security (ICICS), pp. 77- 91, LNCS, Vol 6476, Springer-Verlag, Heidelberg. (2010)
10. De Mello, E.R., Moorsel, A.V., Fraga, J.D.S.: Evaluation of P2P Search Algorithms for Discovering Trust Paths. In: Proc. of 4$^{th}$ European Performance Engineering Conf. on Formal Models and Stochastic Models for Performance Evaluation, pp. 112- 124. (2007)
11. Li, X., Wang, J.: A Global Trust Model of P2P Network Based on Distance-Weighted Recommendation. In: Proc. of IEEE Int. Conf. of Networking, Architecture, and Storage, pp. 281-284. (2009)
12. Crespo, A., Garcia-Molina, H.: Semantic Overlay Networks for P2P Systems. Technical Report, Stanford University. (2002)
13. Schlosser, M.T., Condie, T.E., Kamvar, S.D., Kamvar, A.D.: Simulating a P2P File-Sharing Network. In: Proc. of the 1$^{st}$ Workshop on Semantics in P2P and Grid Computing. (2002)
14. Ganeriwal, S., Srivastava, M.: Reputation-Based Framework for High Integrity Sensor Networks. In: Proc. of the 2$^{nd}$ ACM Workshop on Security of Ad Hoc and Sensor Networks (SAN '04), pp. 66 – 77. (2004)
15. Shafer G.: A Mathematical Theory of Evidence. Princeton University. (1976)
16. Jsang, A.: A Logic for Uncertain Probabilities. In: International Journal of Uncertainty, Fuzziness and Knowledge-Based Systems, 9(3), 279 – 311. (2001)
17. Bloom, B.: Space-Time Trade-Offs in Hash Coding with Allowable Errors. In: Communications of the ACM, 13(7), 422 – 426. (1970)